\documentclass[12pt,prl,amsmath,amssymb,sort&compress,comma,super,showpacs,nofootinbib]{revtex4}

\usepackage{graphics}
\usepackage{epsfig}
\usepackage{subfigure}  

\begin{document}

\newcommand{\etal}{{\it et al.}\/}
\newcommand{\gtwid}{\mathrel{\raise.3ex\hbox{$>$\kern-.75em\lower1ex\hbox{$\sim$}}}}
\newcommand{\ltwid}{\mathrel{\raise.3ex\hbox{$<$\kern-.75em\lower1ex\hbox{$\sim$}}}}
\setcounter{footnote}{0}
\renewcommand{\thefootnote}{\fnsymbol{footnote}}

\title{Contribution of the Electron-Phonon Coupling to the Pairing Interaction in LiFeAs}

\author{
R.A. Jishi$^{1}$ and Douglas Scalapino$^{2}$}
\affiliation{$^1$Department of Physics, California State University,
Los Angeles, CA 90032 USA\\
$^2$Department of Physics, University of California,
Santa Barbara, CA 93106-9530 USA}

\date{\today}

\begin{abstract}
The coupling strengths for intra- and inter-band electron-phonon pair scattering
are calculated for LiFeAs. While the sum of these couplings, which gives the
total electron-phonon coupling $\lambda$, is of order 0.2, we find that their
contribution to the presumed $s^\pm$ superconducting state is reduced by
approximately an order of magnitude.
 
\end{abstract}


\maketitle


Shortly after the discovery of superconductivity in LaFeAsO$_{1-x}$F$_x$,
Boeri \etal\ \cite{ref:1} calculated the electron-phonon coupling $\lambda$
using density-functional perturbation theory. For pure LaFeAsO they found
$\lambda\sim0.21$ and noted that $F$ doping would reduce it. In further
calculations \cite{ref:2} for BaFe$_2$As$_2$, it was found that magnetism
could lead to an enhancement of $\lambda$ by as much as 50\%.  In this work
they found that the largest contributions to the electron-phonon coupling were
concentrated around the $\Gamma(Z)$ and $M$ regions of the Brillouin zone.
If the superconduting gap has an $s^\pm$ structure then the intraband
electron-phonon couplings in the $\Gamma(Z)$ region connect regions of the
Fermi surfaces that have the same sign of the gap and will enhance the pairing
strength. However the couplings around $M$ connect Fermi surface regions on
which the $s^\pm$ gap has different signs and will act to suppress it. If the
small momentum processes dominate, the electron-phonon interaction will enhance
the $s^\pm$ pairing but with an effective coupling $\lambda_{\rm eff}$ which is
smaller than $\lambda$. Recently Ummarino \etal\ \cite{ref:3} have suggested that
the electron-phonon interaction may play an important role in LiFeAs. They argue
that a range of experimental results can be understood using a multiband Eliashberg
model with an interband spin-fluctuation pairing interaction and a large intraband
electron-phonon coupling. Here we discuss density-functional perturbation theory
results for both intra- and inter-band electron-phonon couplings and examine
how they effect the $s^\pm$ pairing strength in LiFeAs. We find that the sum of
these couplings gives $\lambda\sim0.2$, consistent with results for the other
Fe-based materials \cite{ref:1,ref:2}. However, their effective contribution to
the $s^\pm$ pairing strength $\lambda_{\rm eff}$ is reduced by approximately
an order of magnitude due to a cancellation of the intra- and inter-band couplings.

\begin{figure}[htbp]
\includegraphics[height=8.5cm]{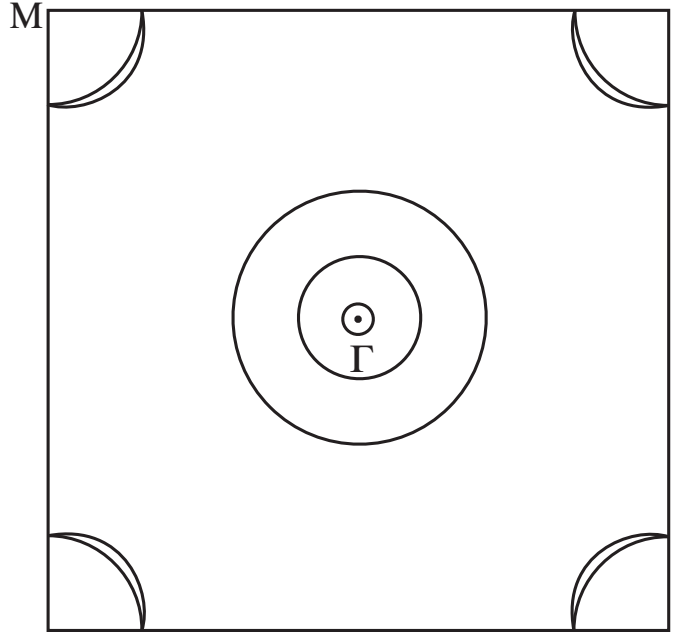}
\caption{A schematic of a cross-section of the LiFeAs Fermi surface in the 2Fe per unit cell
Brillouin zone. Starting from the zone center $\Gamma$ and moving toward M, the bands 
are labeled $\alpha_1$, $\alpha_2$, $\gamma$, $\beta_1$, and $\beta_2$, respectively.\label{fig:1}}
\end{figure}

A schematic of a cross-section of the LiFeAs Fermi surface is shown in Fig.~\ref{fig:1}.
There are three cylindrical hole pockets $\alpha_1$, $\alpha_2$ and $\gamma$
around the $\Gamma$ point and two cylindrical electron pockets $\beta_1$ and
$\beta_2$ around the $M$ point of the 2Fe per unit cell Brillouin zone. One of us (RAJ) previously
carried out calculations of the electronic structure, phonon frequencies and
the electron-phonon coupling $\lambda$ for LiFeAs \cite{ref:4}. Here we are
interested in determining the electron-phonon coupling strength $\lambda(i,j)$
associated with the scattering of electron pairs between the $i$th and $j$th
Fermi surfaces with $i$ and $j$ running over $(\alpha_1,\alpha_2,\gamma,\beta_1,\beta_2)$.
The coupling strength is given by
\begin{equation}
  \lambda(i,j)=\sum_\nu\int\frac{d^3k}{\Omega_{\text{BZ}}}
	                     \int\frac{d^3k'}{\Omega_{\text{BZ}}}\quad
	\frac{2\left|g_\nu\left(ik,jk'\right)\right|^2}{\hbar\omega_{k-k',\nu} N(\varepsilon_F)}\quad
	\delta\left(\varepsilon_{ik}-\varepsilon_F\right) ~
	\delta\left(\varepsilon_{jk'}-\varepsilon_F\right)
\label{eq:1}
\end{equation}
Here $\varepsilon_i(k)$ is the energy of an electron in band $i$ with wave vector
$k$, $\omega_{k-k',\nu}$ is the frequency of a phonon with branch index $\nu$
and crystal momentum $k-k'$, $N(\varepsilon_F)$ is the total single spin
density of states, and $\Omega_{\text{BZ}}$ is the volume of the Brillouin zone. 
The electron-phonon matrix element $g_\nu(ik,jk')$ is
obtained within density functional perturbation theory \cite{ref:5,ref:6}
from the first-order variation of the self-consistent Kohn-Sham \cite{ref:7}
potential $V_{\rm SCF}$ due to atomic displacements,
\begin{equation}
  g_\nu(ik,jk')=\left(\frac{\hbar}{2\omega_{k-k',\nu}}\right)^{1/2}
	\langle\psi_{j,k'}\left|\Delta V_{\rm SCF}\left(k-k'\nu\right)\right|\psi_{i,k}\rangle
\label{eq:2}
\end{equation}
$|\psi_{i,k}\rangle$ is the single-particle Bloch state characterized by wave
vector $k$ and band index $i$, and
\begin{equation}
  \Delta V_{\rm SCF}(q\nu)=\frac{1}{\sqrt N}\sum_ne^{iq\cdot R_n}
	\sum^s_{\alpha=1}\frac{1}{\sqrt{M_\alpha}}\frac{\partial V_{\rm SCF}}{\partial u_{n\alpha}}
	\cdot\varepsilon_\alpha(q\nu)
\label{eq:3}
\end{equation}
is the self-consistent first-order variation of the potential due to atomic 
displacements. In the above
equation, $N$ is the number of unit cells in the crystal, $M_\alpha$ is the mass
of atom $\alpha$ in unit cell $n$, $u_{n\alpha}$ is its displacement, $s$ is
the number of atoms in one unit cell, and $\varepsilon_\alpha(q\nu)=
(\varepsilon_1(q\nu) ~\varepsilon_2(q\nu)\cdots\varepsilon_s(q\nu))$ is the
eigenvector corresponding to the phonon of wave vector $q$ and branch index $\nu$.

The calculation of the electron-phonon coupling constant proceeds in three steps.
First, the electronic energy bands and the density of states are calculated within
density functional theory by carrying out a self-consistent calculation of the
Kohn-Sham equations. Since the electronic properties of the crystal are determined
by the valence electrons, the core electrons are eliminated by using
pseudopotentials for the various atoms in the unit cell. In this work, ultrasoft
pseudopotentials \cite{ref:8} were used; these are computationally efficient. The
wave functions of the valence electrons are expanded in plane waves with a cutoff
energy of 30 Rydbergs, and the charge density is Fourier expanded with an energy
cutoff of 450 Rydbergs. Integration over the Brillouin zone is approximated by a
sum over a set of $k$-points. Since accurate values of the band energies and
density of states are needed for the calculation of the electron-phonon coupling,
a set of 5525 $k$-points in the irreducible Brillouin zone, generated by a
Monkhorst-Pack \cite{ref:9} mesh of $48\times48\times32$ $k$-points, is used.

\begin{figure}[htbp]
\includegraphics[height=8.5cm]{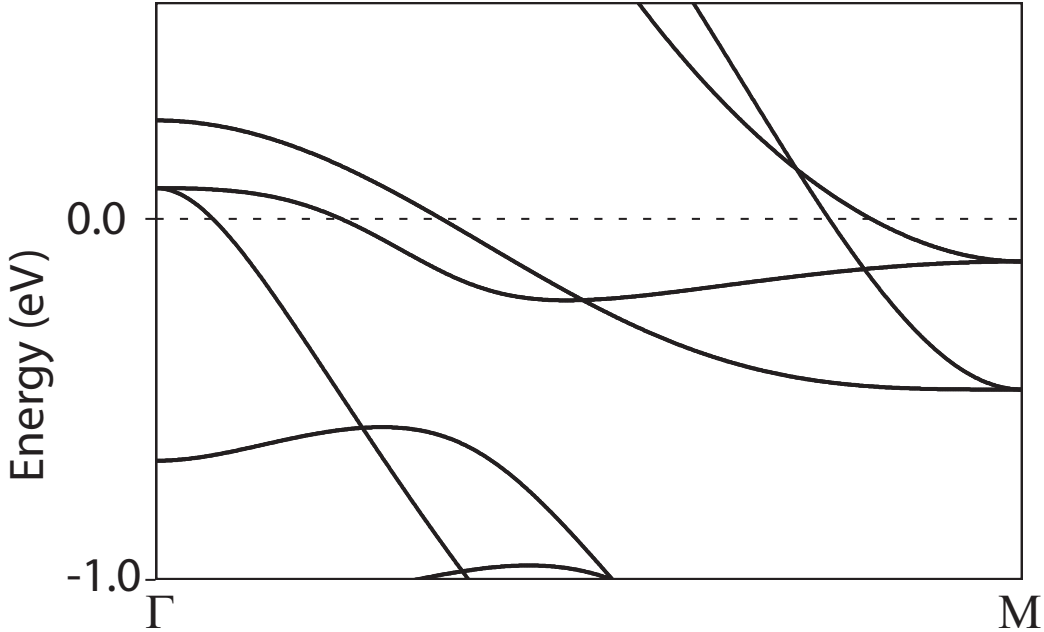}
\caption{Energy bands of LiFeAs along the $\Gamma$M direction. The dashed line 
represents the Fermi energy.
\label{fig:2}}
\end{figure}

In the second step, the phonon frequencies and eigenvectors are calculated within
linear response theory. A self-consistent calculation of the first-order change
in the electron density, brought about by atomic displacements, is carried out.
The change in the density is then used to construct the dynamical matrix, the
diagonalization of which yields the phonon frequencies and eigenvectors. The
calculation is carried out at 140 $q$-points in the irreducible Brillouin zone,
generated by a $12\times12\times8$ mesh of $q$-points.

Finally, the electron-phonon matrix elements $g_\nu(ik,jk')$ and the
coupling strength $\lambda(i,j)$ are calculated. The unit cell of LiFeAs crystal
contains two chemical formulas, and thus there are 28 $(2+16+10)$ valence
electrons in each unit cell. If LiFeAs were an insulator, the lowest 14 bands 
would be occupied, and higher energy bands would be empty. However, LiFeAs 
is not an insulator. In Fig. 2 the calculated energy bands, in the vicinity of
the Fermi energy, are plotted along the $\Gamma$M direction. As this figure shows,
there are three hole pockets at the Brillouin zone
center $\Gamma$ (as well as at $Z(0,0,1/2)$), and two electron pockets at
point $M(1/2,1/2,0)$ (as well as point $A(1/2,1/2,1/2)$). The hole-pockets
result from the bands numbered 12, 13, and 14, while the electron pockets
result from the bands numbered 15 and 16. If LiFeAs were an insulator, bands
12, 13, 14 would be completely filled whereas bands 15 and 16 would be empty.
The hole pockets resulting from
bands 12, 13, and 14 are denoted as the $\alpha_1$, $\alpha_2$, and $\gamma$
pockets, respectively, while the electron pockets resulting from bands 15 and
16 are called the $\beta_1$ and $\beta_2$ pockets, respectively. The electron-phonon
coupling strength $\lambda(i,j)$ corresponding to the scattering of electrons,
by phonon emission or absorption, from band $i$ to band $j$, is obtained using
the Quantum Espresso code \cite{ref:10}, which calculates $g_\nu\left(ik,jk'\right)$
for each phonon mode $(k-k' \nu)$ and electronic energy band indices $i$ and $j$.
As indicated earlier, 
the calculation is carried out for 140 phonon wave vectors in the irreducible
Brillouin zone. Equation 1 is then used to calculate the electron-phonon
coupling strengths $\lambda (i, j)$. In numerical implementation of Eq. (1),
the integration over the Brillouin zone is replaced by a sum over discrete points,
and the Dirac-delta functions are replaced by gaussian functions. The results
for  $\lambda (i, j)$ vary slightly, depending on the width of the gaussian
functions used. In Table I we report the results for  $\lambda (i, j)= \lambda (j, i)$
obtained using gaussian functions of width $0.02$ Rydbergs.
\begin{table}[h]
\caption{The electron-phonon coupling strength $\lambda(i,j)$ for scattering of
pairs between the $i$th and $j$th Fermi surfaces.}
  \begin{tabular}{cccccc}
	  &$\alpha_1$&$\alpha_2$&$\gamma$&$\beta_1$&$\beta_2$\cr
		$\alpha_1$&0.0002&0.0007&0.0012&0.0018&0.0011\cr
		$\alpha_2$&&0.0082&0.0148&0.0138&0.0099\cr
		$\gamma$&&&0.0236&0.0083&0.0060\cr
		$\beta_1$&&&&0.0212&0.0110\cr
		$\beta_2$&&&&&0.0055\cr
  \end{tabular}
\label{tab:1}
\end{table}

We will assume that there is a dominant pairing interaction which gives rise to
a relatively isotropic $s^\pm$ gap and approximate $\Delta_i(k)$ by a constant
$\Delta_i$ on each of the $i$th Fermi surfaces. In this case, the leading order contribution
of the electron-phonon coupling to the pairing strength is given by
\begin{equation}
  \lambda_{\rm eff}=\frac{\displaystyle{\sum_{ij}}\Delta_i\lambda(i,j)\Delta_j}
	{\displaystyle{\sum_i}\frac{N_i(\varepsilon_F)\Delta^2_i}{N(\varepsilon_F)}}
\label{eq:4}
\end{equation}
Here $N_i(\varepsilon_F)$ is the single spin density of states associated with
the $i$th Fermi surface and $N(\varepsilon_F)$ is the total single spin density
of states.

\begin{table}[h]
\caption{Fermi surface single spin density of states and average gap values
with signs appropriate to an $s^\pm$ state.}
  \begin{tabular}{cccccc}
	  &$\alpha_1$&$\alpha_2$&$\gamma$&$\beta_1$&$\beta_2$\cr
		$N_i(\varepsilon_F)(eV)^{-1}$&0.043&0.475&0.665&0.624&0.365\cr
		$\Delta_i$ meV (Ref.~\cite{ref:11})&---&2.5&5.0&$-3.5$&$-3.0$\cr
		$\Delta_i$ meV (Ref.~\cite{ref:12})&---&6&3.4&$-3.4$&$-3.4$\cr
  \end{tabular}
\label{tab:2}
\end{table}

Using the results for $\lambda(i,j)$ given in Table~\ref{tab:1} and the values
of the gaps (Table~\ref{tab:2}) estimated from the ARPES data of Umezawa
\etal\ \cite{ref:11} we find $\lambda_{\rm eff}\simeq0.02$ while from the ARPES
results of Knolle \etal\ \cite{ref:12} we find $\lambda_{\rm eff}\simeq0.03$.
Here we have assumed that the sign of the gap switches between the hole and
electron Fermi surfaces and neglect the contribution of the $\alpha_1$ hole
Fermi surface. In both cases, the electron-phonon interaction
contributes to increasing the $s^\pm$ pairing strength, but due to the cancellation
associated with the sign changing nature of the gap, the contribution is
significantly smaller than the total electron phonon coupling $\lambda$.

\acknowledgments

RAJ gratefully acknowledges support by NSF under grant No. HRD-0932421.
DJS acknowledges the support of the Center for Nanophase Materials Science at ORNL,
which is sponsored by the Division of Scientific User Facilities, U.S.~DOE.

\end{document}